\documentclass{ws-p8-50x6-00}

\begin{document}

\title{Experimental signals of the first phase transition of nuclear matter}

\author{B.~Borderie}

\address{  Institut de Physique Nucl\'eaire, IN2P3-CNRS, F-91406 Orsay Cedex,
 France.~}
 

\maketitle

\abstracts{
Vaporized and multifragmenting sources produced in heavy ion collisions at
intermediate energies are good candidates to investigate the phase diagram
of nuclear matter. The
properties of highly excited nuclear sources which undergo a
simultaneous disassembly into particles are found to sign the presence
of a gas phase. For heavy nuclear sources produced in the Fermi energy domain,
which undergo a simultaneous disassembly into particles and
fragments, a fossil signal (fragment size correlations) reveals the origin
of multifragmentation:spinodal
instabilities which develop in the unstable coexistence region of the phase
diagram of nuclear matter. Studies of fluctuations give a direct signature of
 a first order phase transition through measurements of a negative
 microcanonical heat capacity.
}
\section{Introduction}

The decay of highly excited nuclear systems
through a simultaneous disassembly into fragments and particles, what we call
multifragmentation,
 is a subject of great interest in
nuclear physics.
Indeed multifragmentation should be related to subcritical and/or critical
phenomena. Thus it is fully connected to the nature of the phase transition
which is expected of the liquid-gas type due to the specific form of the
nucleon-nucleon interaction; as van der Waals forces, the nucleon-nucleon
interaction is characterized by attraction at long and intermediate range
and repulsion at short range.

Although multifragmentation has
been observed for many years,
its experimental knowledge 
 was strongly improved only recently with the advent
of powerful devices built in the last decade.
Selecting the ``simplest'' experimental situations, well defined systems
or subsystems which undergo
vaporization (simultaneous disassembly into particles) or multifragmentation
can be thus identified and studied.

It is a difficult task to deduce information on the phase diagram and the
related equation of state
of nuclear matter from nucleus-nucleus collisions at intermediate energies.
 But it is also a very exciting novel physics in relation
with thermodynamics of finite systems (connection to other fields) without
external constraints (pressure,volume)~\cite{GRO,CHOBO}.

\section{From multifragmentation to vaporization : identification of the gas
phase}\label{sec:vapo}
\begin{figure}[htb]
\mbox{\includegraphics[width=8cm]
{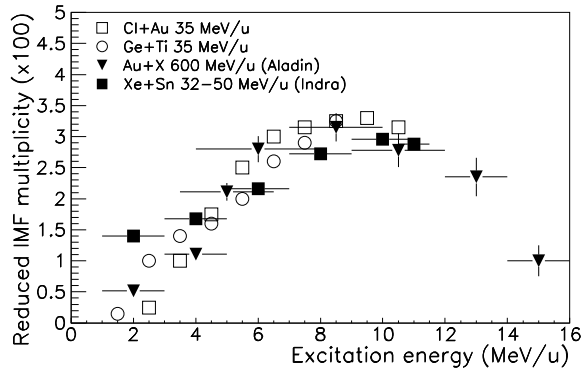}}
\vspace*{-2.4cm} \begin{flushright} \begin{minipage}{3.5cm}
\caption{
Fragment multiplicity (normalised to the number of incident nucleons)
as a function of the excitation energy.
(from~\protect\cite{DU97}). \label{fig:b1}}
\end{minipage} \end{flushright} \end{figure}
Let us first locate in which excitation energy domain
multifragmentation takes place. Fig~\ref{fig:b1} indicates the evolution of
the reduced (normalised to the size of the multifragmenting system) fragment
multiplicity as a function of the excitation energy per nucleon of the system.
A universal behavior characterized by a bell shape curve is observed.
The onset of multifragmentation is observed for excitation energies around 3
MeV per nucleon, the maximum for fragment production is found around 9
MeV per nucleon, i.e. close to the binding energy of nuclei. At higher excitation
energy, the opening of the vaporization channel reduces fragment
production.
\begin{figure}[htb]
\mbox{\includegraphics[width=8cm]
{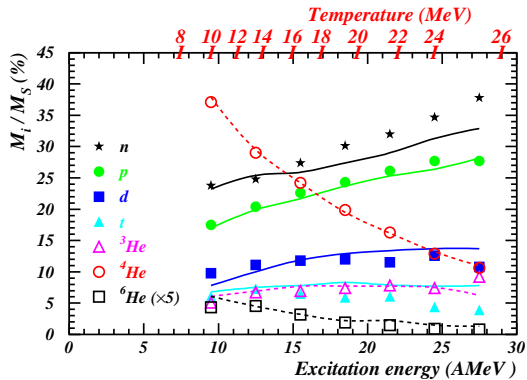}}
\vspace*{-5cm} \begin{flushright} \begin{minipage}{3.5cm}
\caption{
Composition of vaporized quasi-projectiles, formed in 95 AMeV 
$^{36}Ar$+$^{58}Ni$ collisions,
as a function of their excitation
energy per nucleon. Symbols are for
data while the lines (dashed for He isotopes) are the results of the model.
The temperature values used in the model are also given. 
(from~\protect\cite{BO99}). \label{fig:b2}}
\end{minipage} \end{flushright} \end{figure}

The gas phase has been identified by studying the
deexcitation properties of vaporized quasi-projectiles
with A around 36~\cite{BO99}. Chemical composition
(first and second moments) and average kinetic energies of the different
particles are well described by a gas of fermions and bosons
in thermal and chemical equilibrium. Inclusion of a
van der Waals-like behavior (final state excluded volume) was found decisive
to obtain the observed agreement (see for example figure~\ref{fig:b2}).
In the model, the experimental range in
 excitation energy per nucleon of the source was covered
by varying the temperature from 10 to 25 MeV and the free volume 
 was fixed
 at 3$V_{0}$, which corresponds to an average inter-distance between particles
of about 2 fm, close to the range of the nuclear force (freeze-out
configuration).

\section{Thermometry and calorimetry : caloric curves and first-order phase
transition ? }

The plateau observed in the shape of the caloric curve (determined from
calorimetry and nuclear thermometry)
was proposed by the ALADIN collaboration a few years ago as
a signature of a first-order phase transition~\cite{PO95}. Since this observation
works from different collaborations, covering a large range in mass of
nuclei, have been published~\cite{MA97,HA98,KW98,CI00}. Many caloric curves have been obtained which can
roughly be classified in two groups depending on the nuclear thermometer chosen
(isotopic double ratios
using ($^6Li/^7Li$)/($^3He/^4He$) or (d/t)/($^3He/^4He$)). Moreover the
presence of a plateau was not confirmed in these studies, even by the ALADIN
collaboration when looking at properties of target-like spectators in Au+Au
collisions at 1000AMeV~\cite{TR98}. All these studies clearly indicate that no decisive
signal can be extracted. We do not have an absolute nuclear thermometer
and above all, experimentally one does not explore the caloric curve
at constant pressure nor at constant volume. In fact measured caloric curves
are sampling a monodimensional curve on the microcanonical equation of state
surface (T versus energy and volume)~\cite{CHOBO,CHOGA}; for each energy of the system
a different average volume at freeze-out (no more nuclear interaction) is
obtained, depending on the observed partition.

\section{Statistical and dynamical descriptions of multifragmentation}

Many theories have been developed to explain multifragmentation (see for
example ref.~\cite{MO93} for a general review of models). Among the models
some are related to statistical approaches~\cite{GR90,BON95}, valid at and
after freeze-out, whereas others
try to describe the dynamical evolution of systems, from the beginning of
the collision between two nuclei to the fragment formation~\cite{GU96,ONO96}.
I shall very briefly focus here on
two models which will be compared later on to experimental data.
Firstly a statistical description of multifragmentation
(SMM model~\cite{BON95}), in which an
equilibration of a system at low density is assumed. Then the statistical
weight of a given break-up channel $f$, i.e. the number of
microscopic states
leading to this partition, is determined by its entropy
$\Delta \Gamma_f$=$\exp S_f$  
 within the microcanonical framework.
In such an approach the initial parameters as the mass and charge of the
multifragmenting system, its excitation energy, its volume (or density) and
the eventual added radial expansion have to be
backtraced to experimental data.
Secondly, dynamical stochastic mean-field simulations which are obtained by
restoring fluctuations in deterministic one-body kinetic simulations.
In particular in such simulations,
relative to the standard nuclear Boltzmann treatment, an approximate
tool is provided by introducing a noise by means of a brownian force in
the mean field
(Brownian One-Body (BOB) dynamics~\cite{CH94,GU97}).
The magnitude of the force is adjusted to
produce the same growth rate of fluctuations as the full Boltzmann-Langevin
theory~\cite{CHO96}.
Such simulations completely describe the time evolution of the collision and
thus help in learning about nuclear matter and its phase diagram whereas
statistical models start from the phase diagram and have more to do with the
thermodynamical description of finite nuclear systems.

Both descriptions have been successful in reproducing average static and
kinematical properties of fragments (see for example ref.~\cite{TARI00}).
They will be compared in what follows to
more constrained observables
which are expected to bring decisive information on the origin and
properties of multifragmentation.

\section{Correlations in events: spinodal instabilities and equilibration}

\subsection{Fragment size correlations: a fossil signature}

Dynamical simulations predict that during a central collision between heavy
nuclei in the Fermi energy domain (30-40 MeV per nucleon incident energies)
a wide zone of the phase diagram is explored (gentle compression-expansion
cycle) and the fused system enters the liquid-gas coexistence region
 (at low density) and even more precisely the unstable spinodal region
(domain of negative incompressibility). Thus a possible origin of
multifragmentation may be found through the growth of density fluctuations
in this unstable region. Within this theoretical scenario a breakup into
nearly equal-sized ``primitive'' fragments should be favored in relation
with the wave-lengths of the most unstable modes present in the spinodal
region~\cite{AY95}. However this picture is expected to be blurred by
several effects: the beating of differents modes, the presence of large
wave-length instabilities and eventual coalescence of nascent fragments.
Then how to search for a possible ``fossil'' signature of spinodal
decomposition? A few years ago a new method called higher order charge
correlations was proposed in~\cite{MO96}.
All fragments in one event (average fragment charge $<Z>$ and the standard
deviation per event $\triangle$Z) are used to build the charge correlation
for each fragment multiplicity.
\begin{figure}[htb]
\mbox{\includegraphics[width=8cm]
{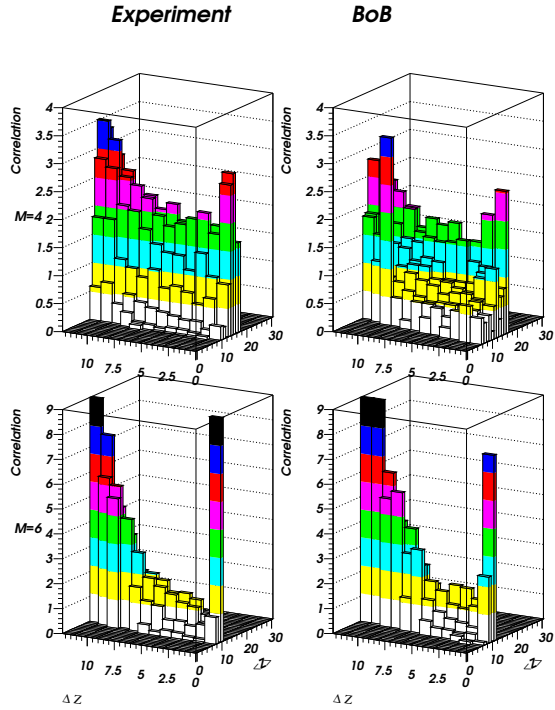}}
\vspace*{-4cm} \begin{flushright} \begin{minipage}{3.5cm}
\caption{
Fragment charge correlations from fused events produced in central collisions
between Xe and Sn at 32 MeV per nucleon incident energy:
comparison between experiment (left) and BOB calculations (right) for
fragment multiplicities equal to 4 and 6.
(from~\protect\cite{TARI00}). \label{fig:b3}}
\end{minipage} \end{flushright} \end{figure}

Figure~\ref{fig:b3} shows results from~\cite{TARI00} for such
correlation functions in experimental
fusion events and BOB simulated events (Xe+Sn system at 32 MeV per nucleon).
For all fragment multiplicities the charge correlation has a peak in the bin
$\Delta Z$ = 0-1, indicating an enhancement of partitions with equal-sized
fragments. This weak but non ambiguous enhancement (0.1\% of events if we
restrict to the bin 0-1 and about 1\% if we enlarge to bin 1-2 to take into
account secondary decays of fragments) is interpreted as a signature
of spinodal instabilities as the origin of multifragmentation in the Fermi
energy domain. Moreover the occurrence of spinodal decomposition signs the
presence of a liquid-gas coexistence region and consequently, although
indirectly, a first order phase transition.

\subsection{Fragment-particle correlations: equilibrium at freeze-out}
Fragment-particle velocity correlations in events have been proposed to experimentally 
measure excitation energy of hot primary fragments produced in
multifragmentation~\cite{MA98}. By means of this technique
multiplicities and relative kinetic energy distributions between fragments and
light charged particles that they evaporate are determined to reconstruct
the excitation energies of fragments.
For the Xe+Sn reaction, the INDRA collaboration has measured the evolution of
the average excitation energy per nucleon
of primary fragments produced in multifragmentation of fused systems at
different incident energies (from 32 to 50 MeV per nucleon)~\cite{HU00}. Within the
error bars a constant value around 3.0-3.5 MeV per nucleon was measured
in good agreement with the approach at equilibrium (SMM). This suggests that
equilibrium is reached at freeze-out. Note that dynamical simulations (BOB)
performed at 32 MeV per nucleon also predict the same excitation of
fragments at freeze-out~\cite{TARI00}.

\textbf{How to reconcile the dynamical (spinodal instabilities) and statistical
(equilibrium at freeze-out) aspects which have been extracted from
correlations ? The following scenario can be proposed:
spinodal instabilities cause multifragmentation but when the system reaches
the freeze-out stage, it has explored enough of the phase space in order to
be describable through an equilibrium approach.}

\section{Kinetic energy fluctuations and negative microcanonical heat
capacity}
\begin{figure}[htb]
\noindent \includegraphics[trim=70 65 247 482,clip,width=6cm]
{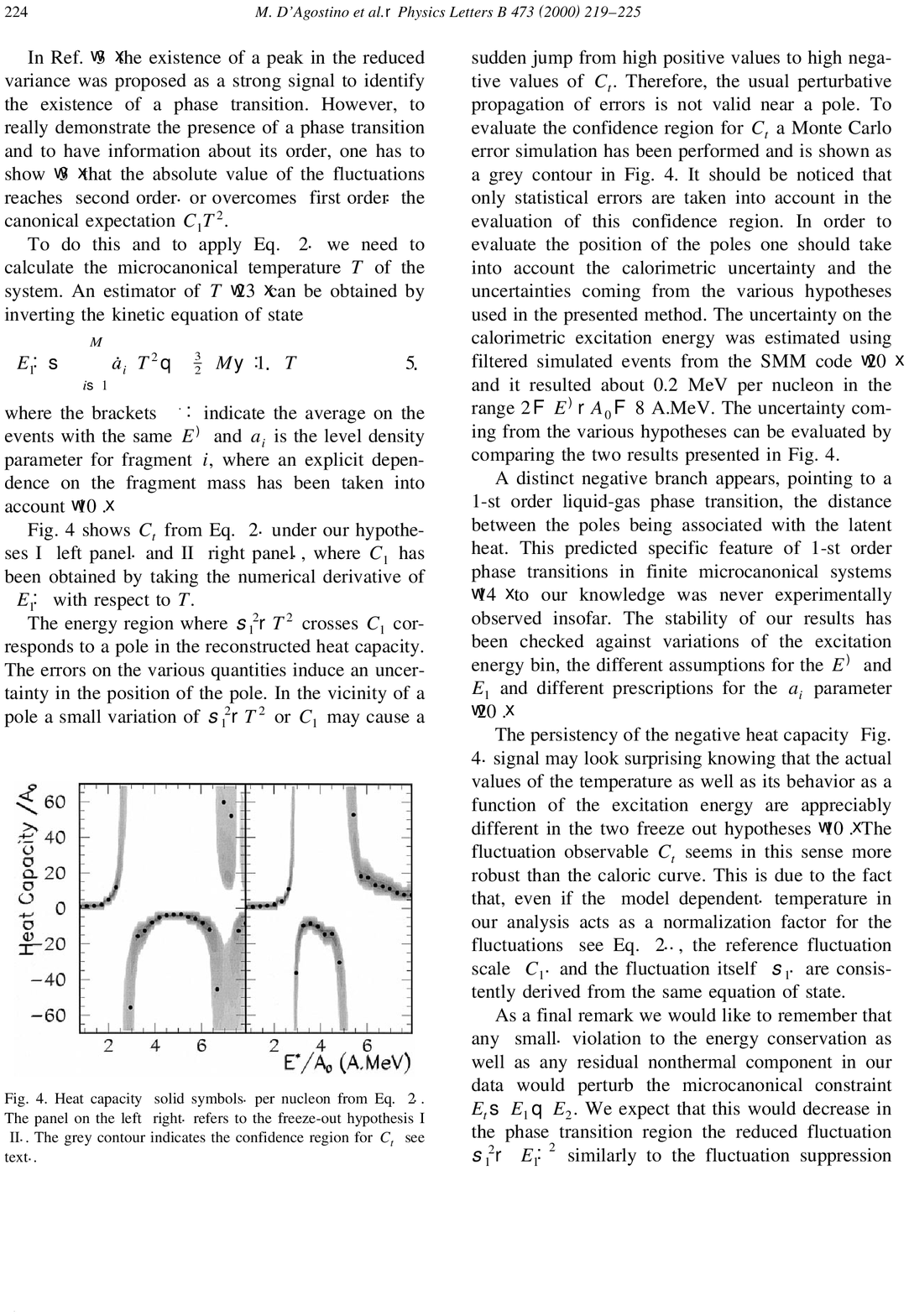}
\begin{flushleft} 
\begin{minipage}{6cm}
\caption{
Measurements of microcanonical heat capacity per nucleon (symbols) as a
function of the excitation energy per nucleon for quasi-projectiles produced
in Au+Au collisions. The two panels refer to
different freeze-out hypotheses. The grey contour indicates the confidence
region for $C_{t}$.
(from~\protect\cite{DA00}). \label{fig:b4}}
\end{minipage} \end{flushleft} 
%\end{figure}

\vspace*{-6.5cm}
%\begin{figure}
\begin{flushright} 
\includegraphics[width=5.5cm]
{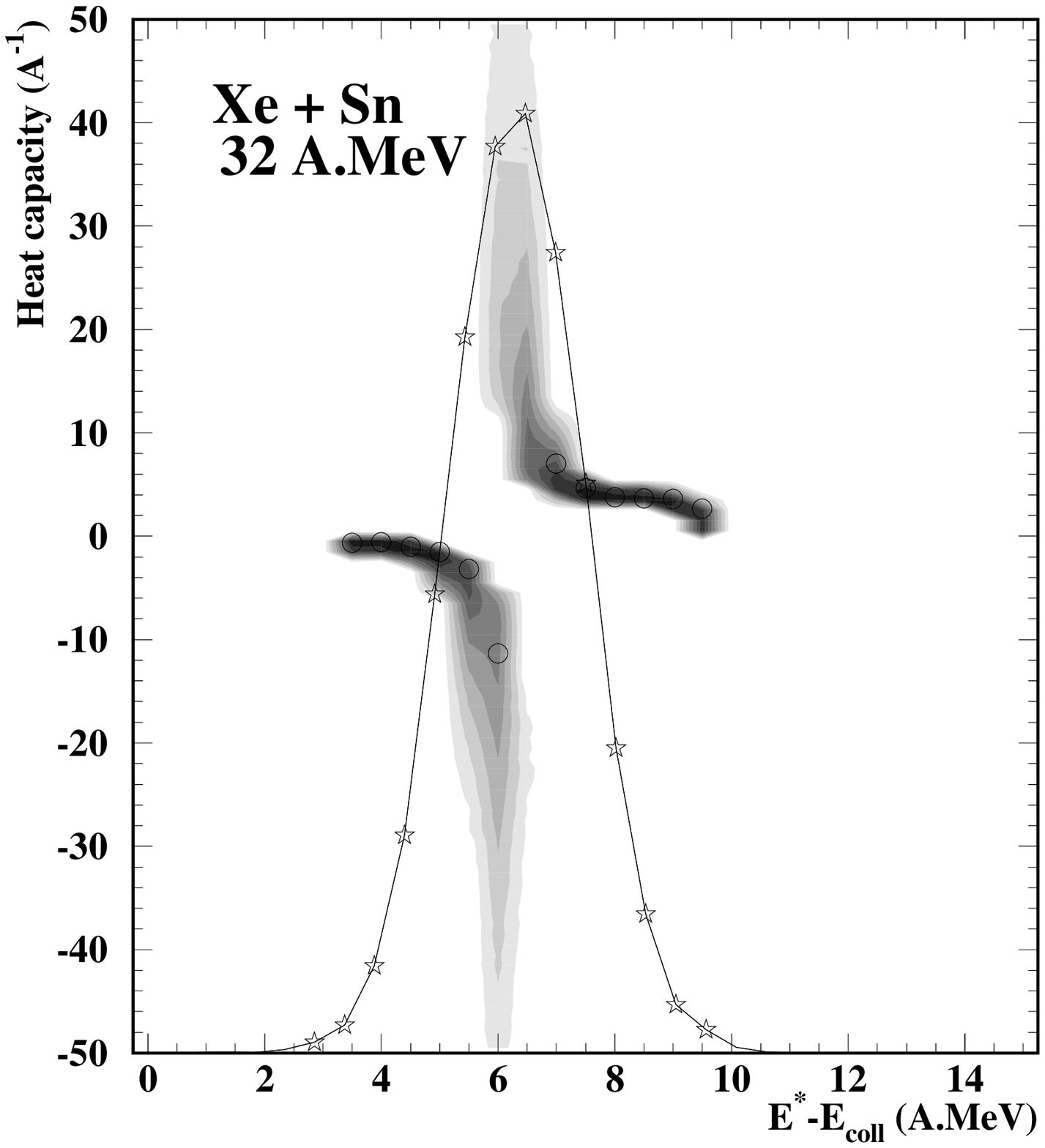}
%\vspace*{-4cm}
\end{flushright}
\begin{flushright}
\begin{minipage}{5.5cm}
\caption{
Same as figure~\ref{fig:b4} for fused nuclei produced in central collisions
between Xe and Sn at 32 MeV per nucleon incident energy.
(from~\protect\cite{LE00}). \label{fig:b5}}
\end{minipage} \end{flushright} \end{figure}
Within the microcanonical equilibrium framework, it was recently
shown~\cite{CHOGU99,CHOBO} that for a given total energy of a system, the
average partial energy stored in a part of the system is a good microcanonical
thermometer, while the associated fluctuations can be
used to construct the heat capacity (see~\cite{CHOBO}). In presence of
a phase transition large
fluctuations are expected to appear as a consequence of the divergence and
of the possible negative branch of the microcanonical heat capacity.
From experiments the most simple decomposition of the total energy $E^{*}$
is in
a kinetic part $E_{1}$ and a potential part $E_{2}$ (Coulomb energy + total
mass excess).
However these quantities have to be determined at freeze-out  and
consequently it is necessary to trace back this
configuration on an event by event basis.
The true configuration needs
the knowledge of all the charged particles evaporated from primary hot
fragments and of the undetected neutrons;
 consequently some reasonable hypotheses have to be done.
Note also that fragment-particle correlations discussed just before can help
to obtain a better knowledge of freeze-out configurations (see~\cite{DABO}).
Then the experimental correlation between the kinetic energy per nucleon
$E_{1}$/$A_{0}$ and
the total excitation energy per nucleon $E^{*}$/$A_{0}$ of the considered
system can be
obtained as well as the reduced variance of the kinetic energy
\mbox{$\sigma_{1}^{2}/<E_{1}^{2}>$}. Finally the
microcanonical temperature of the system can be obtained by inverting the
kinetic equation of state and the total microcanonical heat capacity $C_{t}$
is extracted from the following equations:
\[ {C_1 = \frac{\delta <E_1 / A_0 >}{\delta T}} \; \; \; \; \; \; \;
\rm{and}  \; \; \; \; \; \; \;
 \ {C_t = \frac{C_1^2}{C_1 - \frac{A_0 \sigma_1^2}{T^2}}}\]

Figures~\ref{fig:b4} and \ref{fig:b5} show  results obtained by M.
D'Agostino et al. and the INDRA collaboration for hot nuclei with mass
number around 200 formed in different experimental 
conditions. In figure~\ref{fig:b4}
 the micocanonical heat capacity is calculated
over a large excitation energy range for quasi-projectiles 
(formed in Au+Au collisions at 35 MeV per nucleon incident energy) assuming
two different hypotheses to trace back freeze-out configurations;
figure~\ref{fig:b5} refers to fusion events produced in central Xe+Sn collisions
at 32 MeV per nucleon; in this latter case, a narrower excitation energy
distribution (bell shape curve on the figure) is
 observed. A distinct negative branch is observed, revealing a first
order phase transition. The distances between the poles are associated with the
latent heat. Note that the same location of the pole at high excitation
energy is found
 when similar hypotheses are made for freeze-out reconstruction (left part of 
figure~\ref{fig:b4} and figure~\ref{fig:b5}).
\section{Signatures of critical behavior}
For finite systems, related to the correlation length, a critical region
instead of a critical point is expected.
The following signatures of critical behavior were reported: power laws
have been observed within
selected conditions, critical exponents have been measured in agreement with
with those of a liquid-gas model~\cite{EL98,DA00} assuming that
fragment multiplicity or thermal excitation energy is the control (``order'')
parameter
and
a nuclear scaling function has been evidenced by the
EOS collaboration~\cite{EL98}.
However, as compared to
infinite systems, potential divergences are smoothened over finite regions
of the chosen control parameter and the choice of the fit regions where the
data are assumed to reflect the critical behavior is crucial. Clearly one
needs a precise and objective procedure in order to determine order
parameters and critical signals. Such a methodology was recently proposed for
second order phase transitions~\cite{BO00} and we can expect in the future to dispose
of a similar methodology for first order phase transitions.

\section{Conclusions}
A set of coherent results showing the existence of a first order
phase transition in nuclear matter has
been obtained and the two signals observed related to correlations and
fluctuations constitute a strong starting point for systematic investigations.
Clearly caloric curves do not and can not give a decisive signal of a first
order phase transition.
Selected data have properties compatible with the equilibrium
hypothesis at freeze-out and this framework is up to now chosen for
progressing in the experiment-theory interaction. Experimentally an effort
has to be made to better define configurations at freeze-out, which are
key points to bring more quantitative information (latent heat\ldots ).
On the theoretical side, concerning thermodynamics of finite systems, some
improvements are needed in lattice gas models for example to take into
account the specificities of nuclei (quantal aspects and Coulomb interaction).
Concerning the signatures of critical behavior and the definition of the
critical region, experimentalists need a methodology
dedicated to first order phase transition for finite systems.

\textit{I am highly indebted to my colleagues of the INDRA collaboration and
to R.~Botet, X.~Campi, Ph.~Chomaz, M.~Colonna, M.~D'Agostino and
F.~Gulminelli for valuable discussions.}

\end{document}